\documentclass{article}
\usepackage[utf8]{inputenc}
\usepackage{authblk}
\usepackage{amsmath,amssymb, bm}
\usepackage[breaklinks]{hyperref}
\usepackage{graphicx, float}
\usepackage[square,numbers]{natbib}
\usepackage[thinc]{esdiff}
\hypersetup{
    colorlinks=true,
    linkcolor=blue,
    filecolor=magenta,      
    urlcolor=cyan
    }

\title{The USS Trustee's risky strategy}
\author[1,2]{Neil M Davies}
\author[3]{Jackie Grant}
\author[1,2]{Chin Yang Shapland}

\affil[1]{\textit{MRC Integrative Epidemiology Unit at the University of Bristol, U.K.}}
\affil[2]{\textit{Population Health Sciences, University of Bristol, U.K.}}
\affil[3]{\textit{School of Mathematical and Physical Sciences, Pevensey II, University of Sussex, Falmer, Brighton, U.K.}}

\begin{document}

\maketitle

How much risk, and what types of risk, is the Universities Superannuation Scheme (USS) taking? This is a critical question for universities across the UK and many of their employees. Will the fund have enough money to pay for all our pensions? Will it run out? Or is there a significant risk that we are collectively overpaying? You might have thought that these questions would also be of interest to the USS Trustee, that oversees the fund for members’, or The Pensions Regulator, that regulates it. However, curiously, while the USS has published a plethora of incoherent statistics and opaque analyses, neither it or TPR have bothered to publish estimates of the risk of the scheme defaulting. 

In September 2021, David Miles and James Sefton, Professors of Financial Economics at Imperial College Business School, stepped into this vacuum, publishing \href{https://www.niesr.ac.uk/sites/default/files/publications/NIESR\%20Policy\%20Paper\%20029_0.pdf}{‘How much risk is the USS taking?’} \cite{Miles2021}. The paper presents important, accessible and highly readable analysis which estimates how likely the USS is to default over time. Their work is particularly relevant to the current \href{https://www.ucu.org.uk/article/11767/UCU-higher-education-ballot-timetable-for-action}{UCU dispute with 69 employers} over the benefit cuts that Universities UK (UUK) is planning to implement on the basis of the 2020 USS valuation. 

In this Brief, we assess the assumptions, replicate the results, explore further their model and consider potential extensions. We demonstrate that for a cautious model with reasonable assumptions for assets and asset growth, the fund has a less than 7\% chance of defaulting for the duration that pensions promises are due, but a greater than 80\% chance of being over funded by at least £100bn, and nearly 50\% chance of having over £400bn. 

We offer warm thanks to David Miles and James Sefton for sharing their code and data, for their helpful conversations and clarification. Their analysis is infinitely clearer, better and more credible than anything the USS has produced. We hope this Brief will be the beginning of more work in this area. All errors are our own.

\section*{Exploring the Miles and Sefton model}
Here, we replicate Miles and Sefton’s \cite{Miles2021} results, and assess the plausibility of the assumptions they use given the context of the USS. As with most empirical analysis, their results are highly sensitive to the assumptions used as input to the models, particularly about the mean and variability of rates of return. Nevertheless, their model provides an excellent framework for investigating how different strategies might affect the USS’s risk of default.

Miles and Sefton test whether the funds available on the valuation date can cover all the pensions promised from those assets, as they fall due and without further contribution. Although they state this is “not an assumption that the scheme closes”, the assumption is consistent with a closed scheme and does not model the features of an open scheme. The USS is an open “immature” scheme (i.e. its members are relatively young), so this assumption is highly conservative, as it is likely to increase the risk of default in any given year because in their model the fund will not receive any additional revenue (i.e. no new USS pension members or additional years of accrual), except from the returns from its investments. A second assumption Miles and Sefton make is that the scheme has no “covenant support”. This horrific piece of pension jargon just means that our employers (universities), cannot top up the fund if it looks to be underfunded. This assumption is not realistic, as the USS has perhaps the strongest covenant of any private pension scheme in the country. This is a conservative assumption that likely increases the model’s risk of scheme default, because their model assumes employers cannot provide additional funds to address any deficit. Nevertheless, in our analysis below we make the same assumptions, while acknowledging that they are highly cautious, and we hope to revisit these assumptions in future analysis.

\subsection*{Asset allocation}
The asset allocation refers to the proportion of the fund that is assumed to be invested in either equities, which earn a higher but more variable returns on average, and bonds, which are assumed to have a lower but completely certain return. Miles and Sefton \cite{Miles2021} take the initial assets as March 2020 of £66.5bn, with a portfolio weighted 75\% equities to 25\% bonds and a base case for returns on bonds and find “there is a 40\% probability of the USS running out of funds before all pension promises have been met. About 30\% of the time the USS runs out of assets by 2056.” 

We can reproduce these results. See Figure \ref{fig:MS_Fig2} below, which reproduces Figure 2 of Miles and Sefton \cite{Miles2021}. We also reproduce their results for all figures, including the more optimistic scenarios with reversion to the mean, and zero real rate of return on bonds. We provide code here, and Google sheets to \href{https://docs.google.com/spreadsheets/d/1cKNVyYVEjiQrFOKg2GYckHF-kDPqrW3WU3z8QKQ85wI/edit?usp=sharing}{reproduce Figures 2 and 3} and \href{https://docs.google.com/spreadsheets/d/1IgpQwr7Zjiip_Mi8H_GWea5PpfbWIQT7ruQB500mZsg/edit?usp=sharing}{reproduce Figures 4 and 5} of their paper. R-code is also available on \href{https://github.com/CYShapland/USSBriefs2021}{Github}.

\begin{figure}[ht]
\centering
\includegraphics[scale=0.58]{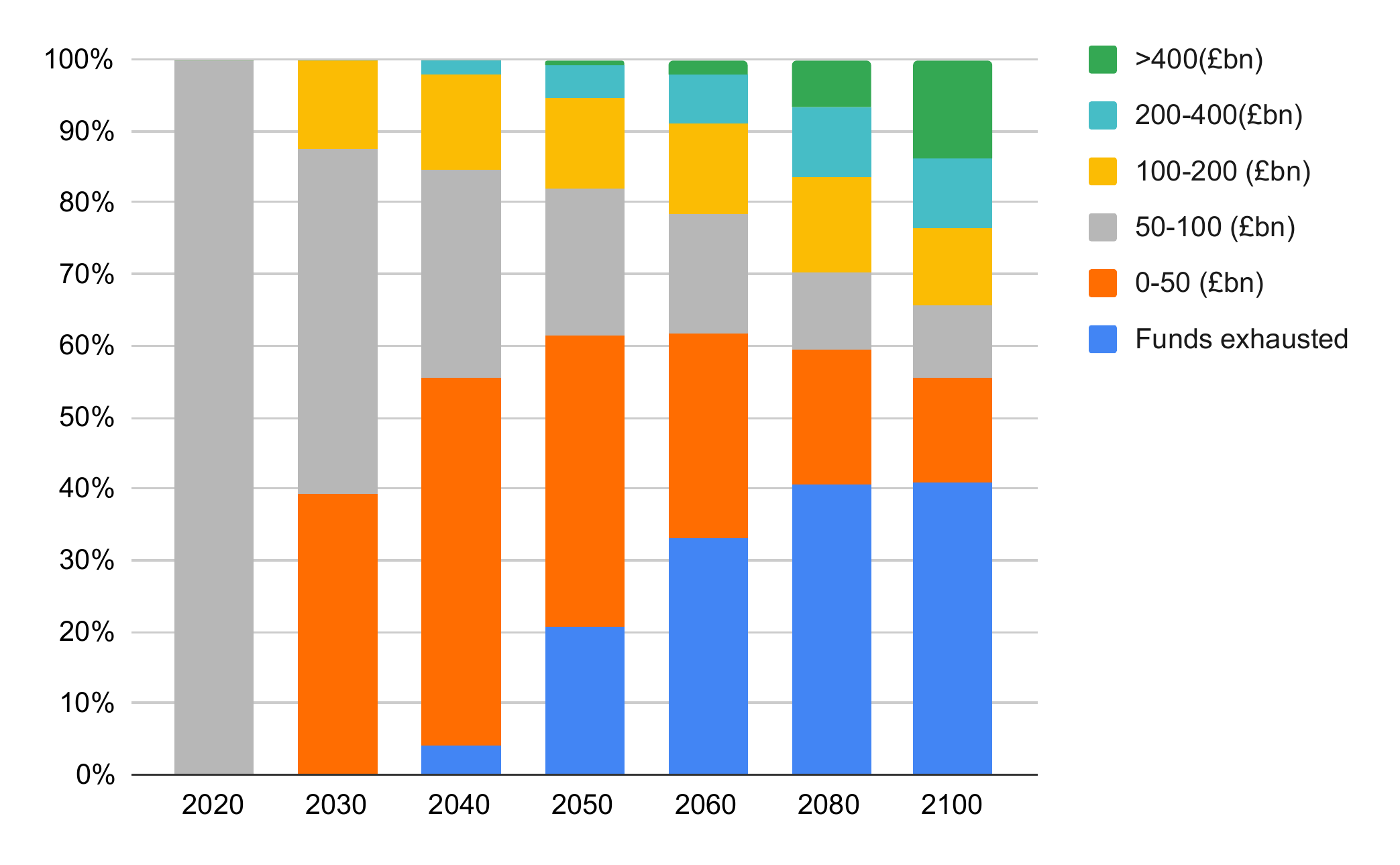}
\caption{Projected distribution of USS fund assets replicating Miles and Sefton Figure 2 \cite{Miles2021}. Starting assets £66.5bn with 75\% allocation to equities, no mean reversion in equity returns and gilt yield from real spot rates as of March 2020, no RPI adjustment, CPI basis. Mean return on equities 4.5\% and standard deviation 17.5\%.}
\label{fig:MS_Fig2}
\end{figure}

\section*{Measuring default risk and asset allocation}
\begin{figure}[ht]
\centering
\includegraphics[scale=0.58]{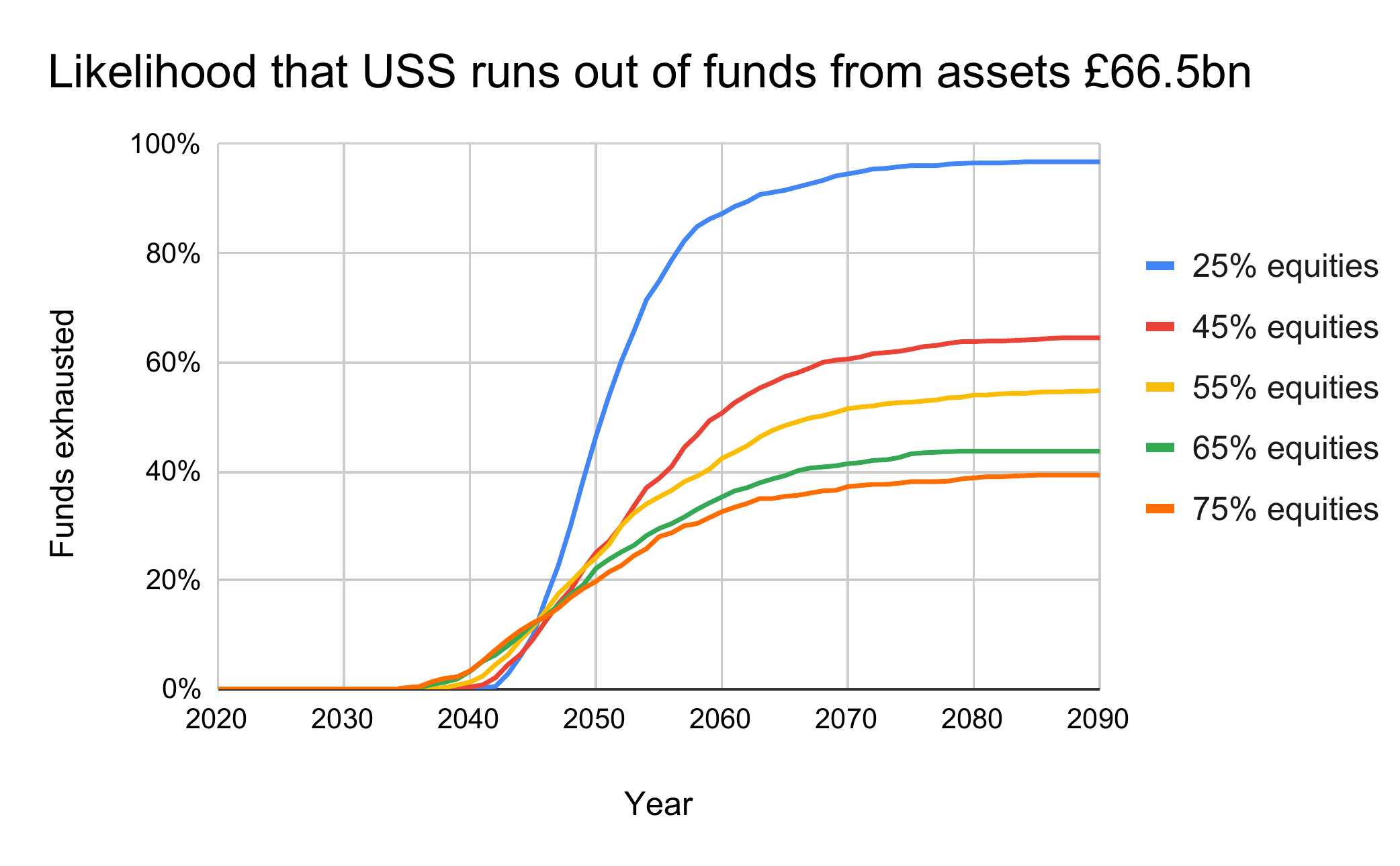}
\caption{Percentage of outcomes where USS funds are exhausted over a range of portfolios from 25\% equities (75\% bonds) to 75\% equities (25\% bonds). Here, as with all figures, the fund is paying promises as they fall due from assets with no contributions from scheme members or employers. CPI basis, using Miles and Sefton \cite{Miles2021} assumptions with no mean reversion, initial assets of £66.5bn, USS cashflows to 2021, bond yield from real spot rates at March 2020 and RPI adjustment 0.5\%. Mean return on equities 4.5\% and standard deviation 17.5\%.}
\label{fig:asset66}
\end{figure}

Next, how does the allocation to equities affect the risk of the scheme defaulting? This risk is presented in Figure \ref{fig:asset66}. In this model we assume the Miles and Sefton \cite{Miles2021} base case values of real yields on index linked gilts as forecast from March 2020. What we see is that as the proportion of the fund that is invested in equities falls, the risk of default substantially increases\footnote{More specifically there is crossover point towards 2050 (in models using the base rate assumptions on bond returns) where the likelihood of default is around 10\% for all asset mixes.}. For example, if the fund is only 25\% invested in equities the risk of default by 2080 is nearly 100\%. The last decade has seen many defined benefit pension funds reduce the amount they invest in company shares (equities), and increase the amount they invest in government bonds. This investment strategy is widely, and perhaps misleadingly, referred to as “derisking”. These models clearly show that the long term risk of default increases as the proportion of the fund invested in equities falls. In the \href{https://en.wikipedia.org/wiki/Nineteen_Eighty-Four}{jargon of USS} “derisking” counter-intuitively increases the risk of default. The \href{https://www.uss.co.uk/about-us/report-and-accounts?search=0333c9f4-f43d-4a05-ab73-130b6644c1e1}{USS 2019 accounts}, p20, report a USS reference portfolio with 60\% equities. The \href{https://www.uss.co.uk/-/media/project/ussmainsite/files/about-us/valuations_yearly/2020-valuation/uss-technical-provisions-consultation-2020-valuation.pdf?rev=89e3e8d0fbb344bf8d9609f6d0eb412e&hash=484A87C8F8D8719BF0AA864D7CC1A3D4}{USS 2020 valuation consultation}, p23, assumes an allocation of 55\% equities for a ‘strong’ covenant and 40\% equities for a ‘tending to strong’ covenant. Increasing the proportion of the fund invested in bonds in this way, is likely to increase the risk of default. These results are consistent with analysis from the \href{https://www.thepensionsregulator.gov.uk/-/media/thepensionsregulator/files/import/pdf/modelling-long-term-funding-objective.ashx}{Government Actuary’s Department} which also suggested that increasing the proportion of DB pension funds invested in bonds is likely to increase the risk of default. Therefore a key message from Miles and Sefton’s \cite{Miles2021} analysis, which is independent of input parameters, is that “derisking” or increasing the proportion of the fund invested in government debt likely increases the risk of default, and is very bad for members and employers.

Miles and Sefton \cite{Miles2021} evaluated the scheme as of 30th March 2020, at the depths of the COVID-19 induced financial panic. \href{https://medium.com/ussbriefs/the-case-for-a-credible-evidence-based-2021-uss-valuation-d5e8c536350c}{Worldwide stock markets} fell by 19\% in the 39 days prior to 30th March 2020. Clearly, since this date, many things have happened. Since March 2020, as can be seen in \href{https://medium.com/ussbriefs/the-case-for-a-credible-evidence-based-2021-uss-valuation-d5e8c536350c}{Figure 1 of USS Briefs 109} the value of worldwide stock markets have increased by 52\%. Even the USS has seen a 33\% increase in its assets with \href{http://www.ussemployers.org.uk/sites/default/files/field/attachemnt/USS\%2031\%20March\%202020\%20valuation\%20-\%20Aon\%20report\%209April2021.pdf}{UUK’s actuary stating} that March 2020 was a ‘poor date for the valuation’ and warned of ‘a danger that too much can be read into the conclusions of a valuation at that date.’ So again, this choice of date is likely to be a conservative assumption that increases the risks of the scheme defaulting, because if its initial assets are lower, then the long term risk of default increases.

\section*{Update to 2021}
\begin{figure}[ht]
\centering
\includegraphics[scale=0.58]{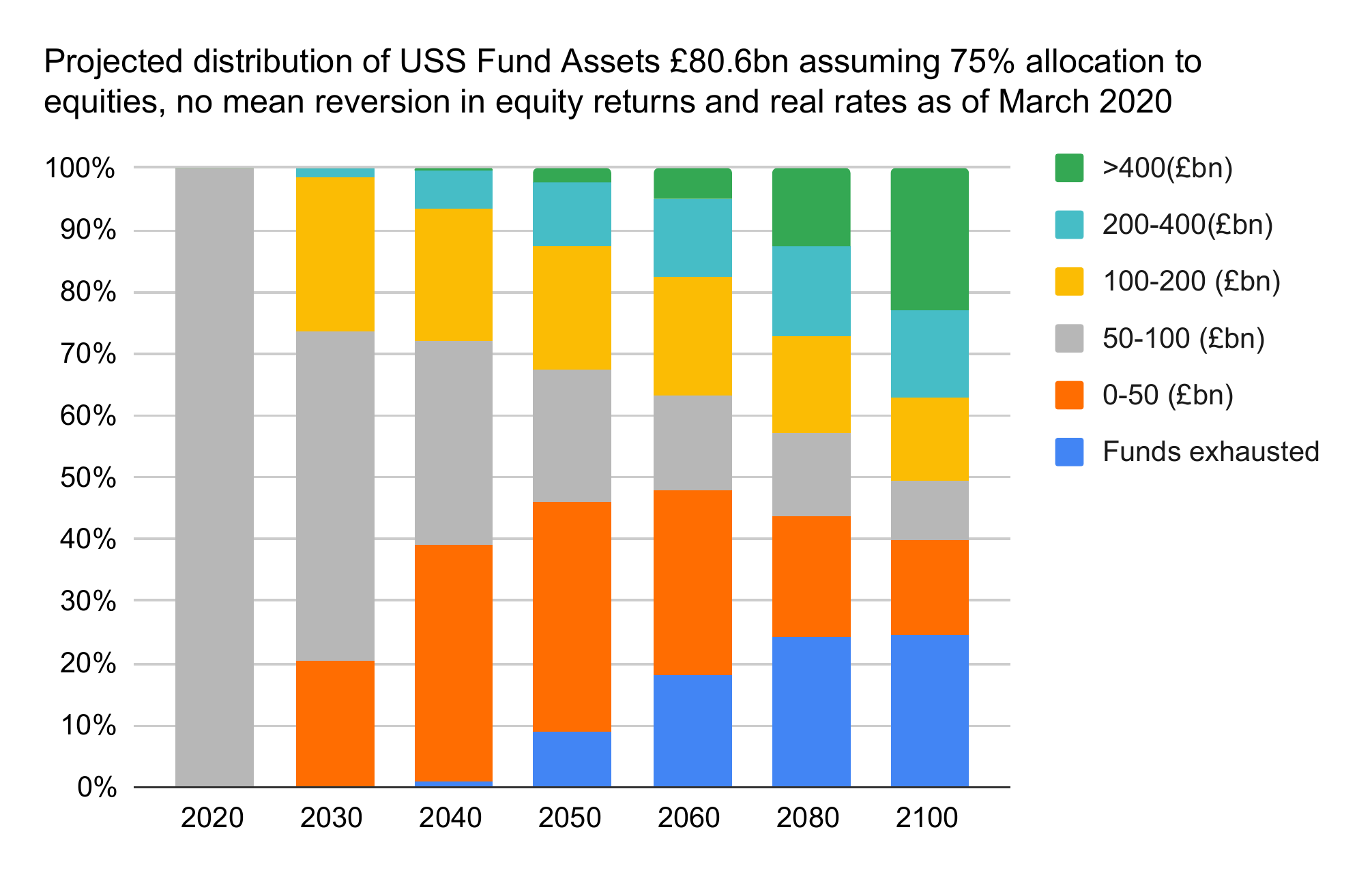}
\caption{Projected distribution of USS funds. Starting asset value of £80.6bn using Miles and Sefton \cite{Miles2021} assumptions of no mean reversion, USS cashflows to 2021, gilt yield from real spot rates at March 2020 and RPI adjustment 0.5\%, CPI basis. Mean return on equities 4.5\% and standard deviation 17.5\%.}
\label{fig:asset80_alloc75}
\end{figure}

 If the initial assets are instead set at £80.6bn (the value of the USS fund at March 2021) the risk of running out of funds before all pension promises have been met falls from over 40\% to 25\%, well within the 67\% prudence level previously used by USS, even under the highly cautious assumptions of this modelling (Figure \ref{fig:asset80_alloc75}). 

On the other hand, the risk of over funding by at least £50bn, for the duration over which pension promises are paid, is always over 50\%. The risk of having over £200bn left in the fund when all promises have been paid is around 35\%, while the risk of having over £400bn left over is over 20\%. Thus, as Miles and Sefton \cite{Miles2021} point out, there are material risks of the scheme running out of money, but there are even larger risks of the scheme having far too much money, and ending up with a very large surplus. Both running out of money and having far too much money are very bad outcomes for members. If the scheme runs out of money, their pensions could be cut, and if the scheme has too much money, they will have paid far too much for their pensions. The employers will also have contributed far too much, and that money could have been used for university investment. 

It is the job of the USS Trustee to balance these risks. Historically pension funds have aimed not for a 50:50 risk of being over or underfunded, but have aimed for a 65:35 risk of being overfunded - this is known as taking a “prudent” approach. It is always possible to decrease the risk of default by increasing the assets in the scheme, but this is expensive, and someone has to pay for it (i.e. active members and their employers). Again, it is the Trustees job to balance the costs of additional prudence on active members, against the benefits of reducing the risk of default. 

\begin{figure}[ht]
\centering
\includegraphics[scale=0.58]{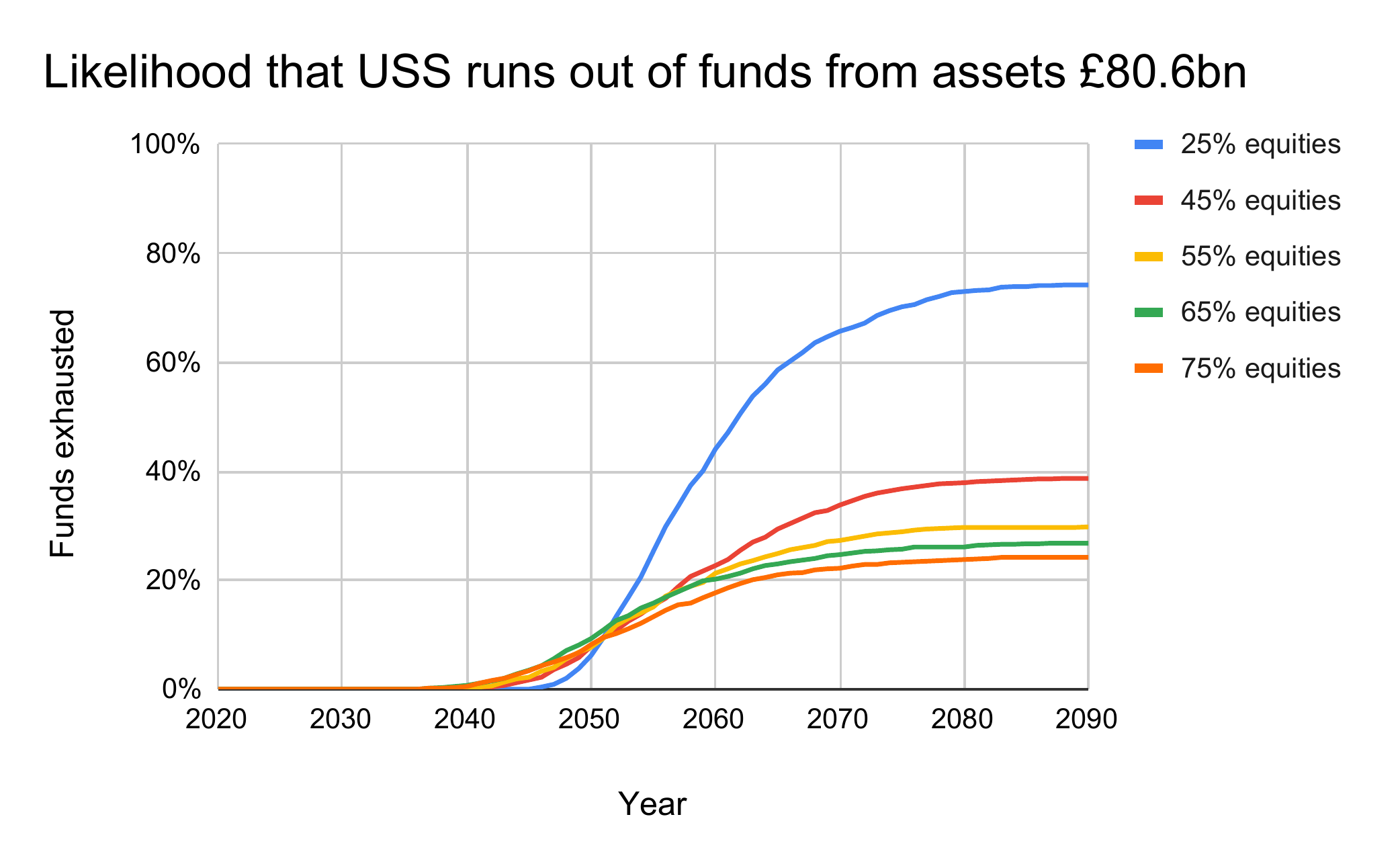}
\caption{Percentage of times USS funds are exhausted over a range of portfolios from 25\% equities (75\% bonds) to 75\% equities (25\% bonds). Here the fund is paying promises as they fall due from assets with no contributions. CPI basis, using Miles and Sefton \cite{Miles2021} assumptions with no mean reversion, initial assets of £80.6bn, USS cashflows to 2021, gilt yield from real spot rates at March 2020 and RPI adjustment 0.5\%. Mean return on equities 4.5\% and standard deviation 17.5\%}
\label{fig:asset80}
\end{figure}

The likelihood of running out of funds across a range of asset allocations, using the value of the USS’s assets at March 2021 is given in Figure \ref{fig:asset80}. Figure \ref{fig:asset80} shows that the risk of default has fallen, as asset values have recovered from the COVID dip. This reduction in the risk of default is particularly large for strategies with higher equities allocations. For example, for asset allocation of 65\% equities there is now only around a 25\% chance the fund will run out of assets before all pension promises have been paid, and a similar 27\% risk of having over £200bn remaining in the fund once all pensions have been promised.

\section*{Evidence based investment returns?}
Finally, how do Miles and Sefton’s \cite{Miles2021} assumptions for investment returns compare to evidence? They assume equities have a real mean return of 4.5\% (SD=17.5\%). These estimates are low compared to estimates of historic returns in the published economics literature. For example, Jord\`a \textit{et al} (2019) \cite{Jorda2019} estimate that equities across 16 developed countries have mean real returns of 6.9\% (SD=21.9\%)\footnote{Further, if we were to weight countries in the Jord\`a \textit{et al.} \cite{Jorda2019} dataset by their GDP, rather than using raw averages, the mean=7.43\% and SD=20.52\%. This would imply even lower rates of default. In addition, these figures implicitly assume that the USS is invested in a single country. Whereas in reality the USS can (and really should), have a globally diversified portfolio to minimise investment risk by investing in a broad range of countries. Using a simple strategy of investing in each country as a proportion of GDP results in returns of equities of mean=7.09 and SD=13.52 (see Figure \ref{fig:asset80_weightedJorda}).}. They are also low compared with the historical real return of all USS investments which is around 7.8\% over the last ten years and 8.1\% over the last five years. Sefton and Miles \cite{Miles2021} justify this by assuming investment costs of between 0.25\% and 0.5\% of assets, and by assuming that future returns will be lower. The cited investment costs are high compared to even retail investments (e.g. index funds can charge as little as 0.03\%), therefore it is unclear why USS costs should be so high\footnote{As an aside USS costs, particularly investment costs, are extraordinarily high, far higher than comparable large schemes reported in the TPR cost analysis. Therefore, there is no reason for USS costs to be as high as Miles and Sefton \cite{Miles2021} assume.}. Miles and Sefton \cite{Miles2021} argue that because the rate on UK government bonds is currently low, we should expect future returns to be lower. However, it is difficult to see how this claim is supported by empirical evidence. For example, Moody’s estimates the 15 year correlation between UK government bonds and worldwide equities to be 0.09, implying around 1\% of the variation in future equities returns can be explained by UK government bond yields (i.e. the yield on government bonds does not meaningfully predict future long-term returns on equities).

Thus, the empirical basis for assuming the mean return on equities of 4.5\%, rather than the just under 7\% seen over the period 1870-2015 is unclear. While it is perfectly reasonable to use assumptions that differ from the historical averages, to be credible, these assumptions, and forecasts need clear empirical support. \href{https://www.uss.co.uk/-/media/project/ussmainsite/files/about-us/valuations_yearly/2020-valuation/march-31-2020-valuation-date.pdf}{USS’s own rationale} for proceeding with a 2020 valuation, on the grounds that a 2021 valuation requires such a significant reduction in return on equities, fails in this respect. Here we aim to explore reasonable assumptions to enlighten and progress the debate, but we look forward to further analysis.

With regard to assumptions on bond returns, we have used Miles and Sefton’s \cite{Miles2021} data throughout. It would be instructive to update bond returns to USS values in 2020 and 2021, and other projections, to better understand the dependency of the modelling on UK government bond yields. The Technical Appendix compares the Bank of England forward curves for March 2020 and 2021 with USS values of March 2020 and Miles and Sefton’s \cite{Miles2021} assumptions for March 2020. Initial analysis suggests broadly similar results for USS 2020 values.  

\begin{figure}[H]
\centering
\includegraphics[scale=0.58]{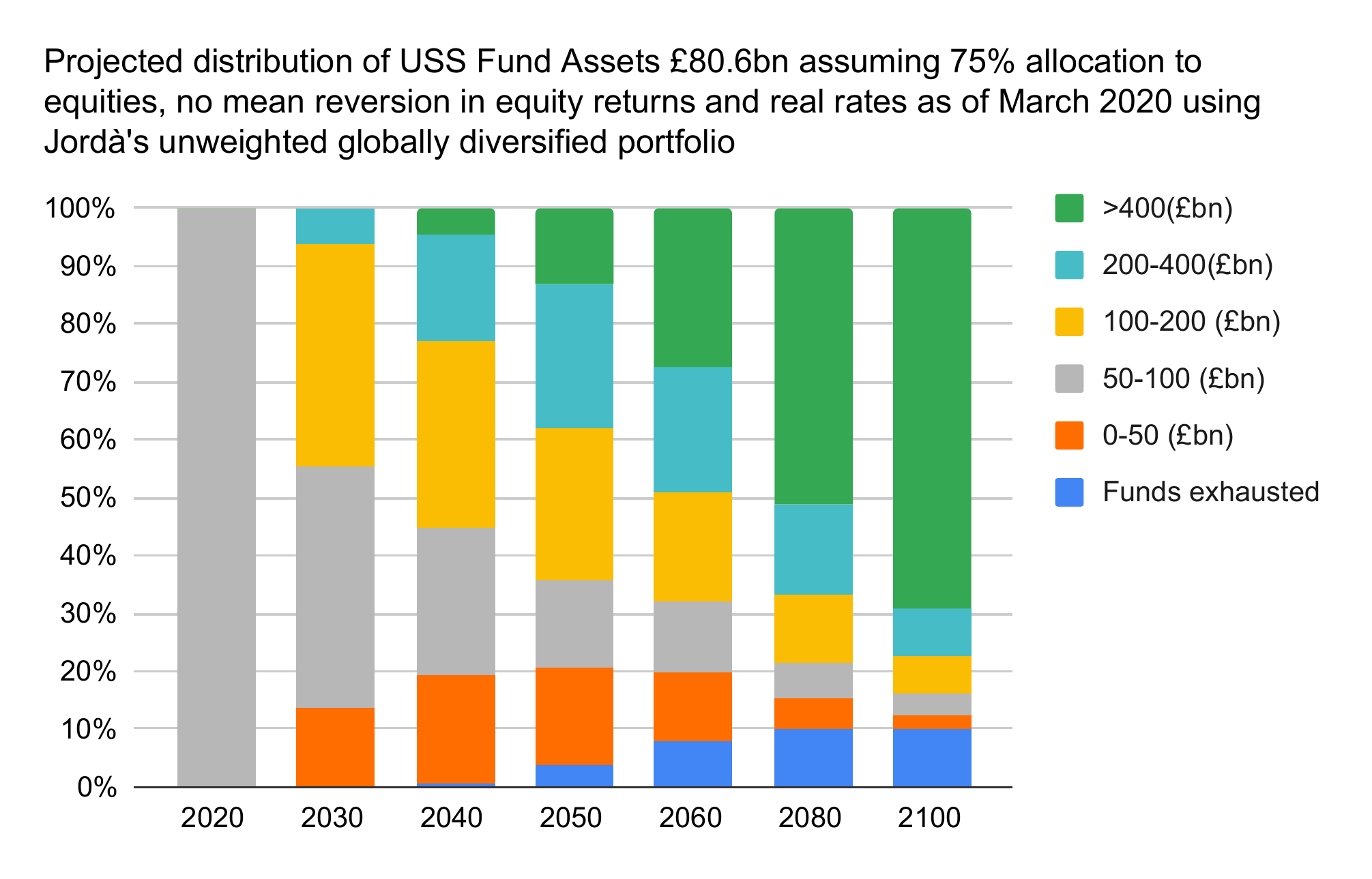}
\caption{Projected distribution of USS funds. Asset value of £80.6bn using Miles and Sefton \cite{Miles2021} assumptions of no mean reversion, USS cashflows to 2021, gilt yield from real spot rates at March 2020 and RPI adjustment 0.5\%, mean return on equities 7.4\% and standard deviation 20.5\% unweighted mean returns on equities for 16 countries 1870-2015 (Jord\`a \textit{et al.} \cite{Jorda2019}).}
\label{fig:asset80_unwieightedJorda}
\end{figure}

\begin{figure}[H]
\centering
\includegraphics[scale=0.58]{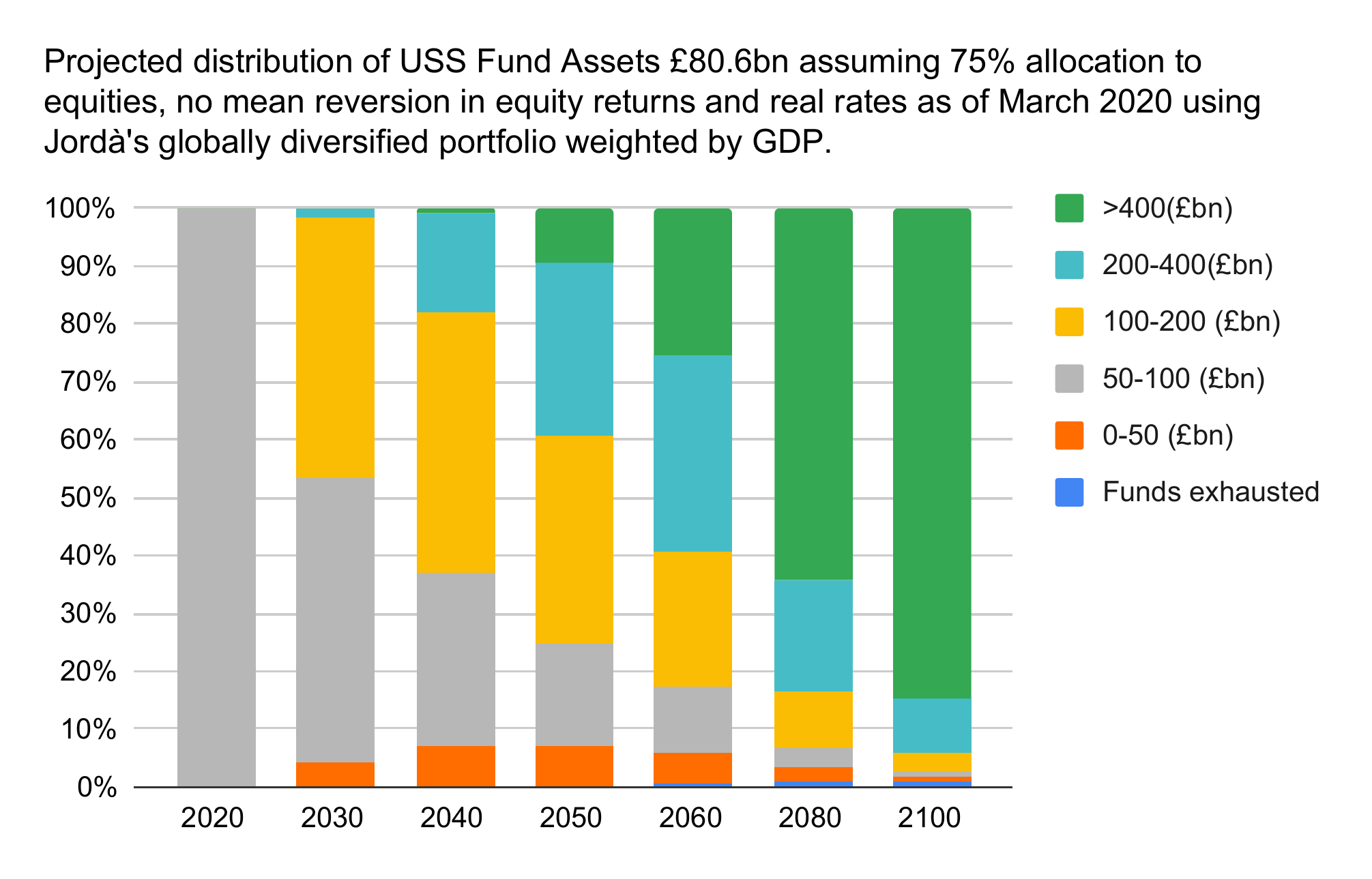}
\caption{Projected distribution of USS funds from asset value of £80.6bn using Miles and Sefton \cite{Miles2021} assumptions of no mean reversion, USS cashflows to 2021, gilt yield from real spot rates at March 2020 and RPI adjustment 0.5\%. Assumed return on equities set to mimic a globally diversified portfolio (Jord\`a \textit{et al.} \cite{Jorda2019} 16 countries 1870-2015), mean 7.1\% and standard deviation 13.5\%.}
\label{fig:asset80_weightedJorda}
\end{figure}

The results when we use estimates from Jord\`a \textit{et al.} (2019) \cite{Jorda2019} for the returns on equities, rather than the more pessimistic asset assumptions used by Miles and Sefton \cite{Miles2021} are presented in Figure 5. The risk of running out of funds before all pension promises have been met falls to around 9\%. However, the risk of over funding by at least £200bn, that is once all pensions promises are paid, is almost 80\%. The risk of having over £400bn left in the fund when all promises have been paid is around 65\%.

It is worth considering other approaches to estimating returns on equities for a fund such as USS. We can model a simple globally diversified portfolio which invests in each country as a proportion of GDP, this results in real returns of equities of mean=7.09\% and SD=13.52\%, from the data in Jord\`a \textit{et al.} (2019) \cite{Jorda2019}. As described in the technical appendix, these values for mean return on equities are not dissimilar to the overall historical performance of the USS fund. This produces the distribution of Figure \ref{fig:asset80_weightedJorda}, where the risk of having insufficient funds to pay all pension promises as they fall due, with no further contributions, is less than 1\% for an equities allocation of 75\%. However the risk of over funding, with such assumptions, by at least £400bn is 85\%. Given that the assumptions of the model are cautious, this risk of overfunding appears unacceptably high.

\begin{figure}[ht]
\centering
\includegraphics[scale=0.58]{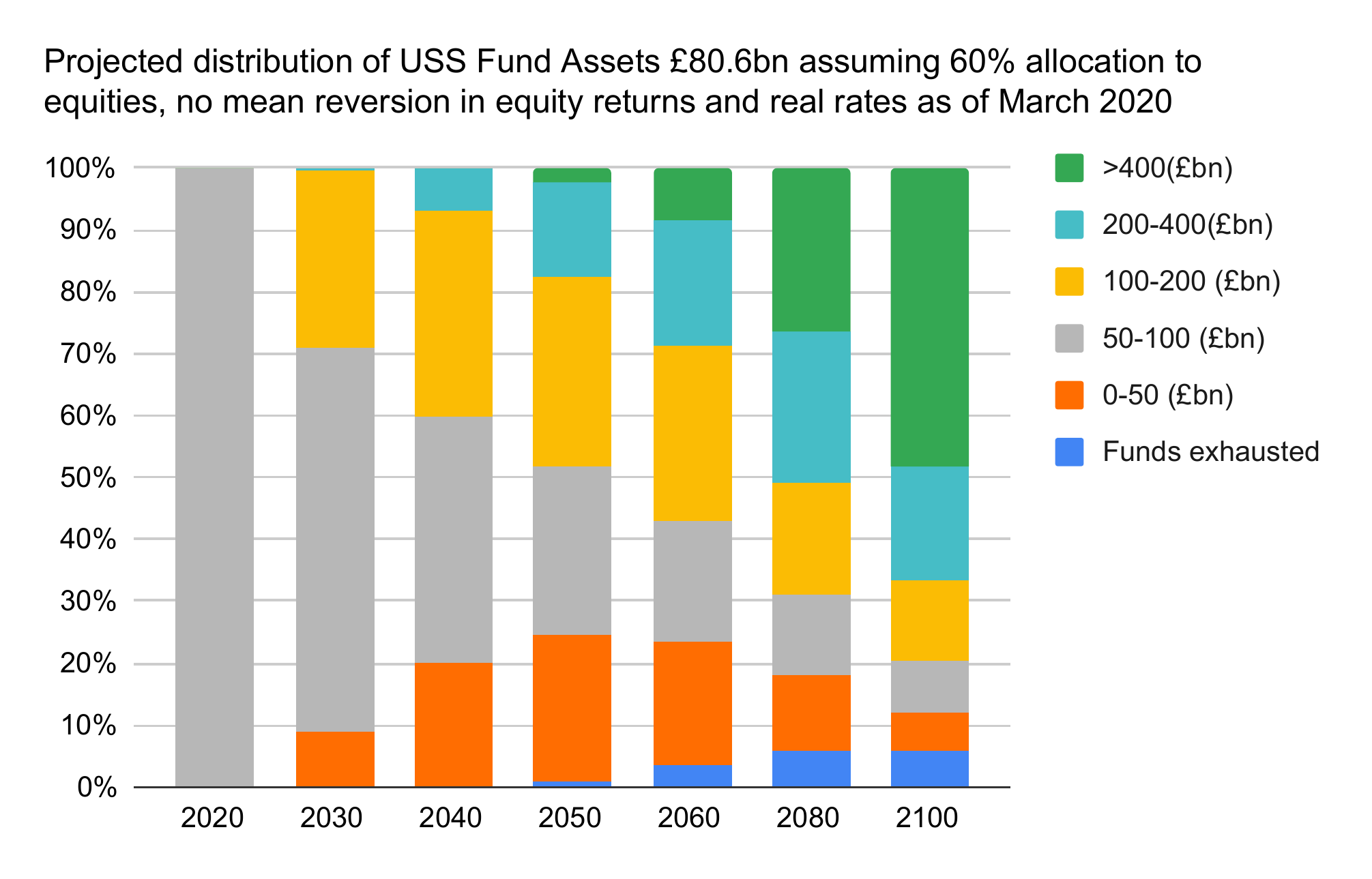}
\caption{Projected distribution of USS funds from asset value of £80.6bn using Miles and Sefton \cite{Miles2021} assumptions of no mean reversion, USS cashflows to 2021, gilt yield from real spot rates at March 2020 and RPI adjustment 0.5\%. Assumed return on equities mean 7.0\% and standard deviation 17\% with a portfolio 60\% equities.}
\label{fig:asset80_USSproj}
\end{figure}

As a final example, Figure \ref{fig:asset80_USSproj}, we consider the USS reference portfolio from the 2019 accounts with 60\% equities, a mean return on equities of 7\%, and a standard deviation of 17\%. The risk of default is a maximum of around 7\% at 2100, a time period that covers the entire duration that pensions promises are due. The risk of default remains less than 5\% up to 2060. The risk of overfunding though is clear, with an 80\% chance of the scheme having over £100bn remaining, and nearly 50\% chance of having over £400bn, having paid all pensions promises under ‘closed scheme’ style assumptions with no covenant. 

\section*{Conclusions: where does this leave us?}
These results demonstrate that even with prudent assumptions:
\begin{itemize}
    \item the risk of default, even over very long periods is low unless the scheme is substantially invested in bonds with a low real return,
    \item the risk of default using up-to-date asset values is even lower, 
    \item the USS’s preferred strategy of increasing the proportion of the fund invested in bonds is likely to increase the risk of default, and 
    \item the 50 year probability of default even using assumptions of a closed scheme is as low as 1\% when using empirically based asset returns and updated USS asset values. This is far lower than conventional levels of prudence used in the pensions industry (i.e. 33\%-35\% as used in the 2014, 2017, and 2018 valuations)
\end{itemize}

Furthermore, even these figures are likely to overestimate the risk of default, as it does not allow for the fact that this scheme is an open immature scheme which has one of the strongest covenants in the UK. It also omits the proactive regulation of open DB schemes in the UK through the TPRs triennial valuations and interventions. It would be relatively straightforward to expand the model to include an intervention every three years that supplied cash top-ups to those paths of the simulation that fell below a given percentage confidence level of a funding target. We leave analysis of possible funding strategies and coherent funding metrics for future work.

Unlike the sum total of analysis published by the USS to date Miles and Sefton’s \cite{Miles2021} analysis and model provides a coherent framework and basis for assessing the risks that members care about - the risk that the scheme will default and the risk that we are all being overcharged. These results, and the results of Miles and Sefton \cite{Miles2021} suggest that we need an urgent detailed, open and reproducible exploration of the USS valuation methodology. This is the only way to restore members' trust after the unnecessary, unjustified, unevidenced and catastrophically pessimistic USS 2020 valuation. This work highlights the significant risk of overfunding, and makes clear that changes are needed to the way risk is quantified in open DB pension schemes. 

It is important to ask what bearing this has on the current valuation and planned cuts to benefits. We remind readers that the \href{https://www.ucu.org.uk/media/11768/USS-trade-dispute-2021/pdf/USS_trade_dispute_2021_Final_002.pdf?utm_source=Lyris&utm_medium=email&utm_campaign=reps&utm_term=broff-jnches-21-22&utm_content=UCU+trade+disputes+with+HE+employers}{UCU formal dispute letter} calls on employers to
\begin{enumerate}
    \item Agree to revoke UUK's proposed cuts to the Defined Benefit pension that were approved by JNC resolution on the 31 August 2021, and not to replace them with any alternative set of changes to benefits and/or increases in member contributions, unless this replacement has the agreement of UCU.
    \item Publicly call on USS to issue an evidence-based, moderately prudent valuation of the financial health of the scheme as at 31 March 2021.
\end{enumerate}

Given this exploration following from the work by Miles and Sefton \cite{Miles2021}, it is clear these demands are not only entirely reasonable, but absolutely necessary. We ask that all employers and scheme members call on UUK to agree to revoke their proposals, and produce a new valuation using a credible, evidence based approach to valuing the scheme's financial health.

\bibliography{USSBriefs2021_DGS}

\newpage

\part*{Technical Appendix for The USS Trustee's risky strategy}

\section*{Introduction}
This document provides the technical appendix for `The USS Trustee’s risky strategy' and structured as follows; Section 2, we describe the Miles and Sefton model. Section 3 is data extraction that includes information on how and where each parameter was extracted to fit Miles and Sefton model. Section 4 summarises the values used for each of the Figure in the main manuscript. The last section consists of the graphical comparison of real forward rate from Miles and Sefton, USS and Bank of England. 

\section*{Miles and Sefton model}

\subsection*{``Risky" assets}
The aim of the Miles and Sefton model is to calculate how assets evolves. Before describing this model we will first define return of risky assets at time $t$, $R_t^e$. $R_t^e$ was log transformed to satisfy the normality assumption, i.e. $r_t^e = log R_t^e$. And assuming distribution to be identically and independently distributed (i.i.d) over time $t$ then we can write returns on risky assets as, 
\begin{equation}
r_t^e= \mu + e_t
\end{equation}
where $\mu$ is the mean log return and $e_t$ is random error with $N(0, \sigma^2_e)$. This model assumes that the returns between $t$ are uncorrelated, which is known as no mean reversion.

For the case where there is correlation between random errors, i.e. presence of mean reversion, the error is then modelled with a moving average process, 
\begin{equation}
e_t = v_t + \sum^q_{i=1} \beta v_{t-i}
\end{equation}
where $v_t \sim N(0, \sigma_v^2)$, $\sigma_v^2$ is the same as $\sigma^2_e$. The idea behind this stochastic model is to captures the change in log returns within a period ($q$) to be similar, then after this time period, the returns can be changed by factor of $\beta$. By setting $\beta<0$ means that a large negative return will be followed by a period of positive return, and vice versa.

\subsection*{``Safe" assets}
Bonds are considered as safe assets, $R_t^f$, i.e. returns from gilts. Miles and Sefton bond yields are calculated from real spot rates at March 2020 and RPI adjustment 0.5\%.

\subsection*{The general model}
Lets assume the proportion of investment portfolio is divided between risky, $\alpha$, and safe assets, $(1-\alpha)$, the initial ($t=0$) asset is denoted as $A_0$ and $p_t$ as pension payments made at time $t$, then the assets, $A_t$, evolves over $t$ can be calculated as;  
\begin{equation}
A_t= (\alpha R_t^e + (1-\alpha)R_t^f)A_{t-1} - p_t
\end{equation}

\section*{Data extraction}
All the data can be found within the Github page: \url{https://github.com/CYShapland/USSBriefs2021}. Bank of England nominal forward rate can be extracted directly from their website (\url{https://www.bankofengland.co.uk/statistics/yield-curves}). Data from Jord\'a \textit{et al.} is publically available from \url{https://dataverse.harvard.edu/file.xhtml?persistentId=doi:10.7910/DVN/GGDQGJ/RDNLLW&version=1.1} 

\subsection*{Real forward rate}
Miles and Sefton used real spot ($R_t^s$) to calculate discount rates ($DR_t$) for time t, then the cumulative $DR_t$ is the returns of the risky assets ($R_t^e$) for each period; 
\begin{align*}
    DR_t = & \left(\dfrac{1+(R_t^s+\delta)}{100}\right)^{t-1} \\
    R_{t-1}^e = & \dfrac{DR_{t-1}}{DR_t - 1} \times 100
\end{align*}
$R_0^e=\delta \times 100$ as $R_0^s$ is 0 at baseline, where $\delta$ is the adjustment factor for converting CPI to RPI.
   
\subsection*{Pension payment}
The pension payment, $p_t$, is calculated as the sum of ``Expected cash flows (£bn) in relation to benefits accrued at 31 March 2020" and ``Expected cash flows (£bn) in relation to benefits projected to be accrued in 2020/21".

\subsection*{Jord\`a's globally diversified portfolio}
Global total returns, real GDP, population size and inflation were extracted from Jord\'a \textit{et al}. First, we calculated the real returns from equity for country i at time t;
\begin{align*}
     r_{i,t} & = \dfrac{1+R_{i,t}}{1+\pi_{i,t}} - 1 
\end{align*}
where $\pi_{i,t}$ is inflation for each country at time t. 
Then real returns is weighted by the proportion of each countries GDP to calculates a global weighted total return;
\begin{align*}
    R_{t} & = \sum^k_{i=1} r_{i,t} \times w_{i,t} \\
    \text{where} \quad w_{i,t} & = \dfrac{N_{i,t}\times rgdp_{i,t}}{\sum^k_{i=1}N_{i,t}\times rgdp_{i,t},}
\end{align*}
population size ($N_{i,t}$) and real GDP per capita ($rgdp_{i,t}$) from each country i at time t. We then use the mean and standard deviation of $R_{t}$ to inform our "risk" assets parameter.

\section*{Parameters summary}
\begin{center}
\begin{table}[ht]
\centering
\caption{Summary of parameters for Figures in the manuscript}
\begin{tabular*}{350pt}{@{\extracolsep\fill}lccccccc@{\extracolsep\fill}}%
    \hline
     \textbf{Figure} & $\bm{A_0}$ \textbf{(£bn)} & $\bm{\alpha}$ & $\bm{\mu}$ & $\bm{\sigma_e^2}$/$\bm{\sigma_v^2}$ & $\bm{\beta}$ & $\bm{q}$ & \textbf{RPI adj.} \\
     \hline
     1 & 66.5 & 0.75 & 0.045 & 0.175 & 0 & 0 & 0 \\
     2 & 66.5 & 0.25-0.75 & 0.045 & 0.175 & 0 & 0 & 0.5 \\
     3 & 80.6 & 0.75 & 0.045 & 0.175 & 0 & 0  & 0.5\\
     4 & 80.6 & 0.25-0.75 & 0.045 & 0.175 & 0 & 0  & 0.5\\
     5 & 80.6 & 0.75 & 0.074 & 0.205 & 0 & 0  & 0.5\\
     6 & 80.6 & 0.75 & 0.071 & 0.135 & 0 & 0  & 0.5\\
     7 & 80.6 & 0.6 & 0.07 & 0.17 & 0 & 0  & 0.5\\

\hline
\end{tabular*}
\end{table}
\end{center}

% JG edits from here 
\newpage

\section*{USS historic returns on assets}

\begin{center}
\begin{table}[ht]
\centering
\caption{Historic returns in percentage on entire USS assets portfolio}
\begin{tabular*}{350pt}{@{\extracolsep\fill}lccccccc@{\extracolsep\fill}}%
    \hline
     \textbf{ } & \multicolumn{1}{c} \textbf{Nominal basis } &  & \multicolumn{1}{c} \textbf{CPI basis} &   \\
     \hline
      
     \textbf{Period} & \textbf{Annual return} & \textbf{Std dev}  & \textbf{Annual return}  & \textbf{Std dev}  \\
     \hline
     1996-2021 & 7.9 & 12.8 & 6.0 & 12.9  \\
     2001-2021 & 5.9 & 13.3 & 3.9 & 13.4 \\
     2011-2021 & 9.8 & 8.3 & 7.8 & 8.6 \\
     2016-2021 & 8.5 & 9.6 & 7.0 & 9.3\\

\hline
\end{tabular*}
\end{table}
\end{center}
Full data set of historic returns at \\ \url{https://docs.google.com/spreadsheets/d/1oFwLQjfv7vPmVAv9BAgnzVp5n5xwuoZyIwi7JV9_biw/} 

%JG edits end here

\newpage

\section*{Figures}

\begin{figure}[ht]
\centering
\includegraphics[scale=0.5]{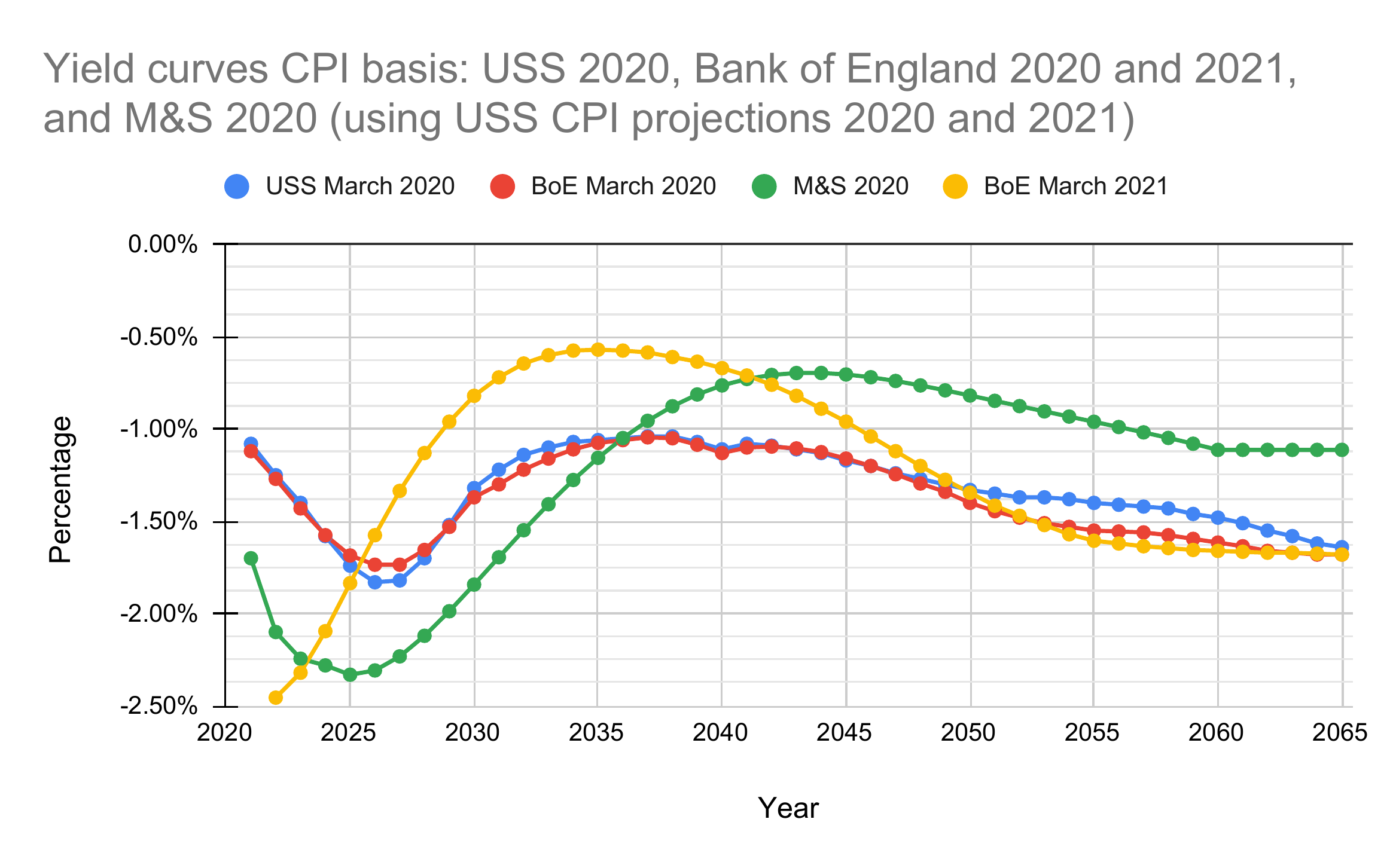}
\caption{Comparison of gilt yield curves used by USS in 2020 and Miles and Sefton versus figures published by the Bank of England for March 2020 and March 2021. For yield curves in 2020, the conversion to CPI basis from nominal basis uses the published USS CPI rates(1-year forward rates as at 31 March 2020). For 2021 USS have only published the Single Equivalent CPI projected value of 2.5\%. Miles and Sefton data using RPI adjustment of 0.5\%.}
\label{fig:Yield_curve}
\end{figure}

\end{document}